
\input epsf
\catcode`\@=11
\font\tensmc=cmcsc10      
\def\smc{\tensmc}

\def\hcorrection#1{\advance\hoffset by #1 }
\def\vcorrection#1{\advance\voffset by #1 }
\def\wlog#1{}
\newif\iftitle@
\outer\def\title{\title@true\vglue 24\p@ plus 12\p@ minus 12\p@
   \bgroup\let\\=\cr\tabskip\centering
   \halign to \hsize\bgroup\tenbf\hfill\ignorespaces##\unskip\hfill\cr}
\def\endtitle{\cr\egroup\egroup\vglue 18\p@ plus 12\p@ minus 6\p@}
\outer\def\author{\iftitle@\vglue -18\p@ plus -12\p@ minus -6\p@\fi\vglue
    12\p@ plus 6\p@ minus 3\p@\bgroup\let\\=\cr\tabskip\centering
    \halign to \hsize\bgroup\smc\hfill\ignorespaces##\unskip\hfill\cr}
\def\endauthor{\cr\egroup\egroup\vglue 18\p@ plus 12\p@ minus 6\p@}
\outer\def\heading{\bigbreak\bgroup\let\\=\cr\tabskip\centering
    \halign to \hsize\bgroup\smc\hfill\ignorespaces##\unskip\hfill\cr}
\def\endheading{\cr\egroup\egroup\nobreak\medskip}

\outer\def\endproclaim{\par\ifdim\lastskip<\medskipamount\removelastskip
  \penalty 55 \fi\medskip\rm}
\outer\def\demo#1{\par\ifdim\lastskip<\smallskipamount\removelastskip
    \smallskip\fi\noindent{\smc\ignorespaces#1\unskip:\enspace}\rm
      \ignorespaces}

\newcount\footmarkcount@
\footmarkcount@=1
\def\makefootnote@#1#2{\insert\footins{\interlinepenalty=100
  \splittopskip=\ht\strutbox \splitmaxdepth=\dp\strutbox
  \floatingpenalty=\@MM
  \leftskip=\z@\rightskip=\z@\spaceskip=\z@\xspaceskip=\z@
  \noindent{#1}\footstrut\rm\ignorespaces #2\strut}}
\def\footnote{\let\@sf=\empty\ifhmode\edef\@sf{\spacefactor
   =\the\spacefactor}\/\fi\futurelet\next\footnote@}
\def\footnote@{\ifx"\next\let\next\footnote@@\else
    \let\next\footnote@@@\fi\next}
\def\footnote@@"#1"#2{#1\@sf\relax\makefootnote@{#1}{#2}}
\def\footnote@@@#1{$^{\number\footmarkcount@}$\makefootnote@
   {$^{\number\footmarkcount@}$}{#1}\global\advance\footmarkcount@ by 1 }

\hyphenation{man-u-script man-u-scripts ap-pen-dix ap-pen-di-ces}
\hyphenation{data-base data-bases}
\ifx\amstexloaded@\relax\catcode`\@=13
  \endinput\else\let\amstexloaded@=\relax\fi
\newlinechar=`\^^J
\def\eat@#1{}
\def\Space@.{\futurelet\Space@\relax}
\Space@. %
\newhelp\athelp@
{Only certain combinations beginning with @ make sense to me.^^J
Perhaps you wanted \string\@\space for a printed @?^^J
I've ignored the character or group after @.}
\def\futureletnextat@{\futurelet\next\at@}
{\catcode`\@=\active
\lccode`\Z=`\@ \lowercase
{\gdef@{\expandafter\csname futureletnextatZ\endcsname}
\expandafter\gdef\csname atZ\endcsname
   {\ifcat\noexpand\next a\def\next{\csname atZZ\endcsname}\else
   \ifcat\noexpand\next0\def\next{\csname atZZ\endcsname}\else
    \def\next{\csname atZZZ\endcsname}\fi\fi\next}
\expandafter\gdef\csname atZZ\endcsname#1{\expandafter
   \ifx\csname #1Zat\endcsname\relax\def\next
     {\errhelp\expandafter=\csname athelpZ\endcsname
      \errmessage{Invalid use of \string@}}\else
       \def\next{\csname #1Zat\endcsname}\fi\next}
\expandafter\gdef\csname atZZZ\endcsname#1{\errhelp
    \expandafter=\csname athelpZ\endcsname
      \errmessage{Invalid use of \string@}}}}
\def\atdef@#1{\expandafter\def\csname #1@at\endcsname}
\newhelp\defahelp@{If you typed \string\define\space cs instead of
\string\define\string\cs\space^^J
I've substituted an inaccessible control sequence so that your^^J
definition will be completed without mixing me up too badly.^^J
If you typed \string\define{\string\cs} the inaccessible control sequence^^J
was defined to be \string\cs, and the rest of your^^J
definition appears as input.}
\newhelp\defbhelp@{I've ignored your definition, because it might^^J
conflict with other uses that are important to me.}
\def\define{\futurelet\next\define@}
\def\define@{\ifcat\noexpand\next\relax
  \def\next{\define@@}%
  \else\errhelp=\defahelp@
  \errmessage{\string\define\space must be followed by a control
     sequence}\def\next{\def\garbage@}\fi\next}
\def\undefined@{}
\def\preloaded@{}
\def\define@@#1{\ifx#1\relax\errhelp=\defbhelp@
   \errmessage{\string#1\space is already defined}\def\next{\def\garbage@}%
   \else\expandafter\ifx\csname\expandafter\eat@\string
         #1@\endcsname\undefined@\errhelp=\defbhelp@
   \errmessage{\string#1\space can't be defined}\def\next{\def\garbage@}%
   \else\expandafter\ifx\csname\expandafter\eat@\string#1\endcsname\relax
     \def\next{\def#1}\else\errhelp=\defbhelp@
     \errmessage{\string#1\space is already defined}\def\next{\def\garbage@}%
      \fi\fi\fi\next}
\def\famzero{\fam\z@}

\def\exp{\mathop{\famzero exp}\nolimits}

\def\lim{\mathop{\famzero lim}}

\def\textfont@#1#2{\def#1{\relax\ifmmode
    \errmessage{Use \string#1\space only in text}\else#2\fi}}
\textfont@\rm\tenrm
\textfont@\it\tenit
\textfont@\sl\tensl
\textfont@\bf\tenbf
\textfont@\smc\tensmc
\let\ic@=\/
\def\/{\unskip\ic@}
\def\textfonti{\the\textfont1 }
\def\t#1#2{{\edef\next{\the\font}\textfonti\accent"7F \next#1#2}}
\let\B=\=
\let\D=\.
\def~{\unskip\nobreak\ \ignorespaces}
{\catcode`\@=\active
\gdef\@{\char'100 }}
\atdef@-{\leavevmode\futurelet\next\athyph@}
\def\athyph@{\ifx\next-\let\next=\athyph@@
  \else\let\next=\athyph@@@\fi\next}
\def\athyph@@@{\hbox{-}}
\def\athyph@@#1{\futurelet\next\athyph@@@@}
\def\athyph@@@@{\if\next-\def\next##1{\hbox{---}}\else
    \def\next{\hbox{--}}\fi\next}
\def\.{.\spacefactor=\@m}
\atdef@.{\null.}
\atdef@,{\null,}
\atdef@;{\null;}
\atdef@:{\null:}
\atdef@?{\null?}
\atdef@!{\null!}
\def\srdr@{\thinspace}
\def\drsr@{\kern.02778em}
\def\sldl@{\kern.02778em}
\def\dlsl@{\thinspace}
\atdef@"{\unskip\futurelet\next\atqq@}
\def\atqq@{\ifx\next\Space@\def\next. {\atqq@@}\else
         \def\next.{\atqq@@}\fi\next.}
\def\atqq@@{\futurelet\next\atqq@@@}
\def\atqq@@@{\ifx\next`\def\next`{\atqql@}\else\def\next'{\atqqr@}\fi\next}
\def\atqql@{\futurelet\next\atqql@@}
\def\atqql@@{\ifx\next`\def\next`{\sldl@``}\else\def\next{\dlsl@`}\fi\next}
\def\atqqr@{\futurelet\next\atqqr@@}
\def\atqqr@@{\ifx\next'\def\next'{\srdr@''}\else\def\next{\drsr@'}\fi\next}

\def\textfontii{\the\textfont2 }
\def\{{\relax\ifmmode\lbrace\else
    {\textfontii f}\spacefactor=\@m\fi}
\def\}{\relax\ifmmode\rbrace\else
    \let\@sf=\empty\ifhmode\edef\@sf{\spacefactor=\the\spacefactor}\fi
      {\textfontii g}\@sf\relax\fi}
\def\nonhmodeerr@#1{\errmessage
     {\string#1\space allowed only within text}}
\def\linebreak{\relax\ifhmode\unskip\break\else
    \nonhmodeerr@\linebreak\fi}
\def\allowlinebreak{\relax
   \ifhmode\allowbreak\else\nonhmodeerr@\allowlinebreak\fi}
\newskip\saveskip@
\def\nolinebreak{\relax\ifhmode\saveskip@=\lastskip\unskip
  \nobreak\ifdim\saveskip@>\z@\hskip\saveskip@\fi
   \else\nonhmodeerr@\nolinebreak\fi}
\def\newline{\relax\ifhmode\null\hfil\break
    \else\nonhmodeerr@\newline\fi}
\def\nonmathaerr@#1{\errmessage
     {\string#1\space is not allowed in display math mode}}
\def\nonmathberr@#1{\errmessage{\string#1\space is allowed only in math mode}}
\def\mathbreak{\relax\ifmmode\ifinner\break\else
   \nonmathaerr@\mathbreak\fi\else\nonmathberr@\mathbreak\fi}
\def\nomathbreak{\relax\ifmmode\ifinner\nobreak\else
    \nonmathaerr@\nomathbreak\fi\else\nonmathberr@\nomathbreak\fi}
\def\allowmathbreak{\relax\ifmmode\ifinner\allowbreak\else
     \nonmathaerr@\allowmathbreak\fi\else\nonmathberr@\allowmathbreak\fi}
\def\pagebreak{\relax\ifmmode
   \ifinner\errmessage{\string\pagebreak\space
     not allowed in non-display math mode}\else\postdisplaypenalty-\@M\fi
   \else\ifvmode\penalty-\@M\else\edef\spacefactor@
       {\spacefactor=\the\spacefactor}\vadjust{\penalty-\@M}\spacefactor@
        \relax\fi\fi}
\def\nopagebreak{\relax\ifmmode
     \ifinner\errmessage{\string\nopagebreak\space
    not allowed in non-display math mode}\else\postdisplaypenalty\@M\fi
    \else\ifvmode\nobreak\else\edef\spacefactor@
        {\spacefactor=\the\spacefactor}\vadjust{\penalty\@M}\spacefactor@
         \relax\fi\fi}
\def\newpage{\relax\ifvmode\vfill\penalty-\@M\else\nonvmodeerr@\newpage\fi}
\def\nonvmodeerr@#1{\errmessage
    {\string#1\space is allowed only between paragraphs}}
\def\smallpagebreak{\relax\ifvmode\smallbreak
      \else\nonvmodeerr@\smallpagebreak\fi}
\def\medpagebreak{\relax\ifvmode\medbreak
       \else\nonvmodeerr@\medpagebreak\fi}
\def\bigpagebreak{\relax\ifvmode\bigbreak
      \else\nonvmodeerr@\bigpagebreak\fi}
\newdimen\captionwidth@
\captionwidth@=\hsize
\advance\captionwidth@ by -1.5in
\def\caption#1{}
\def\topspace#1{\gdef\thespace@{#1}\ifvmode\def\next
    {\futurelet\next\topspace@}\else\def\next{\nonvmodeerr@\topspace}\fi\next}
\def\topspace@{\ifx\next\Space@\def\next. {\futurelet\next\topspace@@}\else
     \def\next.{\futurelet\next\topspace@@}\fi\next.}
\def\topspace@@{\ifx\next\caption\let\next\topspace@@@\else
    \let\next\topspace@@@@\fi\next}
 \def\topspace@@@@{\topinsert\vbox to
       \thespace@{}\endinsert}
\def\topspace@@@\caption#1{\topinsert\vbox to
    \thespace@{}\nobreak
      \smallskip
    \setbox\z@=\hbox{\noindent\ignorespaces#1\unskip}%
   \ifdim\wd\z@>\captionwidth@
   \centerline{\vbox{\hsize=\captionwidth@\noindent\ignorespaces#1\unskip}}%
   \else\centerline{\box\z@}\fi\endinsert}
\def\midspace#1{\gdef\thespace@{#1}\ifvmode\def\next
    {\futurelet\next\midspace@}\else\def\next{\nonvmodeerr@\midspace}\fi\next}
\def\midspace@{\ifx\next\Space@\def\next. {\futurelet\next\midspace@@}\else
     \def\next.{\futurelet\next\midspace@@}\fi\next.}
\def\midspace@@{\ifx\next\caption\let\next\midspace@@@\else
    \let\next\midspace@@@@\fi\next}
 \def\midspace@@@@{\midinsert\vbox to
       \thespace@{}\endinsert}
\def\midspace@@@\caption#1{\midinsert\vbox to
    \thespace@{}\nobreak
      \smallskip
      \setbox\z@=\hbox{\noindent\ignorespaces#1\unskip}%
      \ifdim\wd\z@>\captionwidth@
    \centerline{\vbox{\hsize=\captionwidth@\noindent\ignorespaces#1\unskip}}%
    \else\centerline{\box\z@}\fi\endinsert}
\mathchardef\prime@="0230
\def\prime{{{}\prime@{}}}
\def\prim@s{\prime@\futurelet\next\pr@m@s}

\def\,{\relax\ifmmode\mskip\thinmuskip\else\thinspace\fi}
\def\!{\relax\ifmmode\mskip-\thinmuskip\else\negthinspace\fi}
\def\frac#1#2{{#1\over#2}}

\def\:{\nobreak\hskip.1111em{:}\hskip.3333em plus .0555em\relax}
\def\intic@{\mathchoice{\hskip5\p@}{\hskip4\p@}{\hskip4\p@}{\hskip4\p@}}
\def\negintic@
 {\mathchoice{\hskip-5\p@}{\hskip-4\p@}{\hskip-4\p@}{\hskip-4\p@}}
\def\intkern@{\mathchoice{\!\!\!}{\!\!}{\!\!}{\!\!}}
\def\intdots@{\mathchoice{\cdots}{{\cdotp}\mkern1.5mu
    {\cdotp}\mkern1.5mu{\cdotp}}{{\cdotp}\mkern1mu{\cdotp}\mkern1mu
      {\cdotp}}{{\cdotp}\mkern1mu{\cdotp}\mkern1mu{\cdotp}}}
\newcount\intno@
\def\iint{\intno@=\tw@\futurelet\next\ints@}
\def\iiint{\intno@=\thr@@\futurelet\next\ints@}
\def\iiiint{\intno@=4 \futurelet\next\ints@}
\def\idotsint{\intno@=\z@\futurelet\next\ints@}
\def\ints@{\findlimits@\ints@@}
\newif\iflimtoken@
\newif\iflimits@
\def\findlimits@{\limtoken@false\limits@false\ifx\next\limits
 \limtoken@true\limits@true\else\ifx\next\nolimits\limtoken@true\limits@false
    \fi\fi}
\def\multintlimits@{\intop\ifnum\intno@=\z@\intdots@
  \else\intkern@\fi
    \ifnum\intno@>\tw@\intop\intkern@\fi
     \ifnum\intno@>\thr@@\intop\intkern@\fi\intop}
\def\multint@{\int\ifnum\intno@=\z@\intdots@\else\intkern@\fi
   \ifnum\intno@>\tw@\int\intkern@\fi
    \ifnum\intno@>\thr@@\int\intkern@\fi\int}
\def\ints@@{\iflimtoken@\def\ints@@@{\iflimits@
   \negintic@\mathop{\intic@\multintlimits@}\limits\else
    \multint@\nolimits\fi\eat@}\else
     \def\ints@@@{\multint@\nolimits}\fi\ints@@@}
\def\Sb{_\bgroup\vspace@
        \baselineskip=\fontdimen10 \scriptfont\tw@
        \advance\baselineskip by \fontdimen12 \scriptfont\tw@
        \lineskip=\thr@@\fontdimen8 \scriptfont\thr@@
        \lineskiplimit=\thr@@\fontdimen8 \scriptfont\thr@@
        \Let@\vbox\bgroup\halign\bgroup \hfil$\scriptstyle
            {##}$\hfil\cr}
\def\endSb{\crcr\egroup\egroup\egroup}
\def\Sp{^\bgroup\vspace@
        \baselineskip=\fontdimen10 \scriptfont\tw@
        \advance\baselineskip by \fontdimen12 \scriptfont\tw@
        \lineskip=\thr@@\fontdimen8 \scriptfont\thr@@
        \lineskiplimit=\thr@@\fontdimen8 \scriptfont\thr@@
        \Let@\vbox\bgroup\halign\bgroup \hfil$\scriptstyle
            {##}$\hfil\cr}
\def\endSp{\crcr\egroup\egroup\egroup}
\def\Let@{\relax\iffalse{\fi\let\\=\cr\iffalse}\fi}
\def\vspace@{\def\vspace##1{\noalign{\vskip##1 }}}
\def\aligned{\,\vcenter\bgroup\vspace@\Let@\openup\jot\m@th\ialign
  \bgroup \strut\hfil$\displaystyle{##}$&$\displaystyle{{}##}$\hfil\crcr}
\def\endaligned{\crcr\egroup\egroup}
\def\matrix{\,\vcenter\bgroup\Let@\vspace@
    \normalbaselines
  \m@th\ialign\bgroup\hfil$##$\hfil&&\quad\hfil$##$\hfil\crcr
    \mathstrut\crcr\noalign{\kern-\baselineskip}}
\def\endmatrix{\crcr\mathstrut\crcr\noalign{\kern-\baselineskip}\egroup
                \egroup\,}
\newtoks\hashtoks@
\hashtoks@={#}
\def\format{\crcr\egroup\iffalse{\fi\ifnum`}=0 \fi\format@}
\def\format@#1\\{\def\preamble@{#1}%
  \def\c{\hfil$\the\hashtoks@$\hfil}%
  \def\r{\hfil$\the\hashtoks@$}%
  \def\l{$\the\hashtoks@$\hfil}%
  \setbox\z@=\hbox{\xdef\Preamble@{\preamble@}}\ifnum`{=0 \fi\iffalse}\fi
   \ialign\bgroup\span\Preamble@\crcr}

\def\cases{\left\{\,\vcenter\bgroup\vspace@
     \normalbaselines\openup\jot\m@th
       \Let@\ialign\bgroup$##$\hfil&\quad$##$\hfil\crcr
      \mathstrut\crcr\noalign{\kern-\baselineskip}}

\newif\iftagsleft@
\tagsleft@true
\def\TagsOnRight{\global\tagsleft@false}
\def\tag#1$${\iftagsleft@\leqno\else\eqno\fi
 \hbox{\def\pagebreak{\global\postdisplaypenalty-\@M}%
 \def\nopagebreak{\global\postdisplaypenalty\@M}\rm(#1\unskip)}%
  $$\postdisplaypenalty\z@\ignorespaces}
\interdisplaylinepenalty=\@M
\def\allowdisplaybreak@{\def\allowdisplaybreak{\noalign{\allowbreak}}}
\def\displaybreak@{\def\displaybreak{\noalign{\break}}}
\def\align#1\endalign{\def\tag{&}\vspace@\allowdisplaybreak@\displaybreak@
  \iftagsleft@\lalign@#1\endalign\else
   \ralign@#1\endalign\fi}
\def\ralign@#1\endalign{\displ@y\Let@\tabskip\centering\halign to\displaywidth
     {\hfil$\displaystyle{##}$\tabskip=\z@&$\displaystyle{{}##}$\hfil
       \tabskip=\centering&\llap{\hbox{(\rm##\unskip)}}\tabskip\z@\crcr
             #1\crcr}}
\def\lalign@
 #1\endalign{\displ@y\Let@\tabskip\centering\halign to \displaywidth
   {\hfil$\displaystyle{##}$\tabskip=\z@&$\displaystyle{{}##}$\hfil
   \tabskip=\centering&\kern-\displaywidth
        \rlap{\hbox{(\rm##\unskip)}}\tabskip=\displaywidth\crcr
               #1\crcr}}
\def\overrightarrow{\mathpalette\overrightarrow@}
\def\overrightarrow@#1#2{\vbox{\ialign{$##$\cr
    #1{-}\mkern-6mu\cleaders\hbox{$#1\mkern-2mu{-}\mkern-2mu$}\hfill
     \mkern-6mu{\to}\cr
     \noalign{\kern -1\p@\nointerlineskip}
     \hfil#1#2\hfil\cr}}}
\def\overleftarrow{\mathpalette\overleftarrow@}
\def\overleftarrow@#1#2{\vbox{\ialign{$##$\cr
     #1{\leftarrow}\mkern-6mu\cleaders\hbox{$#1\mkern-2mu{-}\mkern-2mu$}\hfill
      \mkern-6mu{-}\cr
     \noalign{\kern -1\p@\nointerlineskip}
     \hfil#1#2\hfil\cr}}}
\def\overleftrightarrow{\mathpalette\overleftrightarrow@}
\def\overleftrightarrow@#1#2{\vbox{\ialign{$##$\cr
     #1{\leftarrow}\mkern-6mu\cleaders\hbox{$#1\mkern-2mu{-}\mkern-2mu$}\hfill
       \mkern-6mu{\to}\cr
    \noalign{\kern -1\p@\nointerlineskip}
      \hfil#1#2\hfil\cr}}}
\def\underrightarrow{\mathpalette\underrightarrow@}
\def\underrightarrow@#1#2{\vtop{\ialign{$##$\cr
    \hfil#1#2\hfil\cr
     \noalign{\kern -1\p@\nointerlineskip}
    #1{-}\mkern-6mu\cleaders\hbox{$#1\mkern-2mu{-}\mkern-2mu$}\hfill
     \mkern-6mu{\to}\cr}}}
\def\underleftarrow{\mathpalette\underleftarrow@}
\def\underleftarrow@#1#2{\vtop{\ialign{$##$\cr
     \hfil#1#2\hfil\cr
     \noalign{\kern -1\p@\nointerlineskip}
     #1{\leftarrow}\mkern-6mu\cleaders\hbox{$#1\mkern-2mu{-}\mkern-2mu$}\hfill
      \mkern-6mu{-}\cr}}}
\def\underleftrightarrow{\mathpalette\underleftrightarrow@}
\def\underleftrightarrow@#1#2{\vtop{\ialign{$##$\cr
      \hfil#1#2\hfil\cr
    \noalign{\kern -1\p@\nointerlineskip}
     #1{\leftarrow}\mkern-6mu\cleaders\hbox{$#1\mkern-2mu{-}\mkern-2mu$}\hfill
       \mkern-6mu{\to}\cr}}}
\def\sqrt#1{\radical"270370 {#1}}
\def\dots{\relax\ifmmode\let\next=\ldots\else\let\next=\tdots@\fi\next}
\def\tdots@{\unskip\ \tdots@@}
\def\tdots@@{\futurelet\next\tdots@@@}
\def\tdots@@@{$\mathinner{\ldotp\ldotp\ldotp}\,
   \ifx\next,$\else
   \ifx\next.\,$\else
   \ifx\next;\,$\else
   \ifx\next:\,$\else
   \ifx\next?\,$\else
   \ifx\next!\,$\else
   $ \fi\fi\fi\fi\fi\fi}
\def\text{\relax\ifmmode\let\next=\text@\else\let\next=\text@@\fi\next}
\def\text@@#1{\hbox{#1}}
\def\text@#1{\mathchoice
 {\hbox{\everymath{\displaystyle}\def\textfonti{\the\textfont1 }%
    \def\textfontii{\the\textfont2 }\textdef@@ T#1}}
 {\hbox{\everymath{\textstyle}\def\textfonti{\the\textfont1 }%
    \def\textfontii{\the\textfont2 }\textdef@@ T#1}}
 {\hbox{\everymath{\scriptstyle}\def\textfonti{\the\scriptfont1 }%
   \def\textfontii{\the\scriptfont2 }\textdef@@ S\rm#1}}
 {\hbox{\everymath{\scriptscriptstyle}\def\textfonti{\the\scriptscriptfont1 }%
   \def\textfontii{\the\scriptscriptfont2 }\textdef@@ s\rm#1}}}
\def\textdef@@#1{\textdef@#1\rm \textdef@#1\bf
   \textdef@#1\sl \textdef@#1\it}

\def\textdef@#1#2{\def\next{\csname\expandafter\eat@\string#2fam\endcsname}%
\if S#1\edef#2{\the\scriptfont\next\relax}%
 \else\if s#1\edef#2{\the\scriptscriptfont\next\relax}%
 \else\edef#2{\the\textfont\next\relax}\fi\fi}
\scriptfont\itfam=\tenit \scriptscriptfont\itfam=\tenit
\scriptfont\slfam=\tensl \scriptscriptfont\slfam=\tensl
\mathcode`\0="0030
\mathcode`\1="0031
\mathcode`\2="0032
\mathcode`\3="0033
\mathcode`\4="0034
\mathcode`\5="0035
\mathcode`\6="0036
\mathcode`\7="0037
\mathcode`\8="0038
\mathcode`\9="0039
\def\Cal{\relax\ifmmode\let\next=\Cal@\else
     \def\next{\errmessage{Use \string\Cal\space only in math mode}}\fi\next}
\def\Cal@#1{{\fam2 #1}}
\def\bold{\relax\ifmmode\let\next=\bold@\else
   \def\next{\errmessage{Use \string\bold\space only in math
      mode}}\fi\next}\def\bold@#1{{\fam\bffam #1}}
\mathchardef\Gamma="0000
\mathchardef\Delta="0001
\mathchardef\Theta="0002
\mathchardef\Lambda="0003
\mathchardef\Xi="0004
\mathchardef\Pi="0005
\mathchardef\Sigma="0006
\mathchardef\Upsilon="0007
\mathchardef\Phi="0008
\mathchardef\Psi="0009
\mathchardef\Omega="000A
\mathchardef\varGamma="0100
\mathchardef\varDelta="0101
\mathchardef\varTheta="0102
\mathchardef\varLambda="0103
\mathchardef\varXi="0104
\mathchardef\varPi="0105
\mathchardef\varSigma="0106
\mathchardef\varUpsilon="0107
\mathchardef\varPhi="0108
\mathchardef\varPsi="0109
\mathchardef\varOmega="010A
\def\fontlist@{\\{\tenrm}\\{\sevenrm}\\{\fiverm}\\{\teni}\\{\seveni}%
 \\{\fivei}\\{\tensy}\\{\sevensy}\\{\fivesy}\\{\tenex}\\{\tenbf}\\{\sevenbf}%
 \\{\fivebf}\\{\tensl}\\{\tenit}\\{\tensmc}}
\def\dodummy@{{\def\\##1{\global\let##1=\dummyft@}\fontlist@}}
\newif\ifsyntax@
\newcount\countxviii@
\def\newtoks@{\alloc@5\toks\toksdef\@cclvi}
\def\nopages@{\output={\setbox\z@=\box\@cclv \deadcycles=\z@}\newtoks@\output}
\def\syntax{\syntax@true\dodummy@\countxviii@=\count18
\loop \ifnum\countxviii@ > \z@ \textfont\countxviii@=\dummyft@
   \scriptfont\countxviii@=\dummyft@ \scriptscriptfont\countxviii@=\dummyft@
     \advance\countxviii@ by-\@ne\repeat
\dummyft@\tracinglostchars=\z@
  \nopages@\frenchspacing\hbadness=\@M}
\def\magstep#1{\ifcase#1 1000\or
 1200\or 1440\or 1728\or 2074\or 2488\or
 \errmessage{\string\magstep\space only works up to 5}\fi\relax}
{\lccode`\2=`\p \lccode`\3=`\t
 \lowercase{\gdef\tru@#123{#1truept}}}

\def\scaletype#1{\mag=#1\relax
 \hsize=\expandafter\tru@\the\hsize
 \vsize=\expandafter\tru@\the\vsize
 \dimen\footins=\expandafter\tru@\the\dimen\footins}

\def\scalefont#1#2\andcallit#3{\edef\font@{\the\font}#1\font#3=
  \fontname\font\space scaled #2\relax\font@}
\def\Mag@#1#2{\ifdim#1<1pt\multiply#1 #2\relax\divide#1 1000 \else
  \ifdim#1<10pt\divide#1 10 \multiply#1 #2\relax\divide#1 100\else
  \divide#1 100 \multiply#1 #2\relax\divide#1 10 \fi\fi}
\def\scalelinespacing#1{\Mag@\baselineskip{#1}\Mag@\lineskip{#1}%
  \Mag@\lineskiplimit{#1}}
\def\wlog#1{\immediate\write-1{#1}}
\catcode`\@=\active
%
%
%
%
%
%

\headline={\ifnum\pageno=1\firstheadline\else
\ifodd\pageno\rightheadline \else\leftheadline\fi\fi}
\def\firstheadline{\hfil}
\def\rightheadline{\hfil}
\def\leftheadline{\hfil}
	\footline={\ifnum\pageno=1\firstfootline\else\otherfootline\fi}
\def\firstfootline{\rm\hss\folio\hss}
\def\otherfootline{\hfil}

 1
 1
 1

scaled\magstephalf

scaled\magstephalf

scaled\magstephalf

scaled\magstephalf

\font\tenbf=cmbx10
\font\tenrm=cmr10
\font\tenit=cmti10

\font\eightrm=cmr8
\font\eightit=cmti8

\parindent=1.2pc
\magnification=\magstep1
\hsize=6.0truein
\vsize=8.6truein
\nopagenumbers

\centerline{\tenbf DEGENERATE BESS}
\baselineskip=18pt
\centerline{\tenbf AT FUTURE $e^+e^-$ COLLIDERS
\footnote"$^*$"{\eightrm\baselineskip=10pt
Talk given at the Workshop on Physics and Experiments
with Linear Colliders, September 8-12, 1995, Morioka-Appi, Japan.
This work was partially supported by a grant under the
European Human Mobility Program on "Tests of Electroweak
Symmetry Breaking at Future European Colliders".}}
\baselineskip=13pt
\baselineskip=13pt


\centerline{\eightrm DANIELE DOMINICI}
\baselineskip=12pt
\centerline{\eightit Dip. di Fisica, Univ.
di Firenze, Sezione I.N.F.N. Firenze}
\baselineskip=10pt
\centerline{\eightit  Lgo E. Fermi 2, 50125 Firenze, Italy}
\centerline{\eightrm E-mail: dominici\@fi.infn.it}\vglue0.2cm

\vglue0.6cm
\centerline{\eightrm ABSTRACT}
\vglue0.2cm
{\rightskip=3pc
 \leftskip=3pc
 \eightrm\baselineskip=10pt\noindent
An effective lagrangian describing a strong interacting
electroweak sector is considered. It contains new vector and axial-vector
resonances all degenerate in mass and mixed with $W$ and $Z$.
The model, for large mass of these degenerate gauge bosons, becomes
identical to the standard model in the limit of infinite Higgs mass.
The limits
on the parameter space of this model from future $e^+e^-$
colliders are
presented.
\vglue0.6cm}
\def\lmu{{{{\tilde L}}_\mu}}

\def\rmu{{{ \tilde R}_\mu}}

\def\gs{{g''}}

\def\dmu{{\partial_\mu}}

\def\gp{g'}
\def\gpt{{{g}^\prime}}
\def\gptd{ g^{\prime 2}}

\def\eps{{\epsilon}}
\def\tr{{ {tr}}}

\def\f{\frac}
\def\Wt{\tilde{{W}}}

\def\Yt{\tilde{{Y}}}
\def\tW{\tilde W}
\def\tY{\tilde Y}
\def\tL{\tilde L}
\def\tR{\tilde R}
\def\L{{\cal L}}
\def\s{s_\theta}
\def\c{c_\theta}

\def\de{\partial}
\def\eps{\epsilon}
\def\be{$$}
\def\ee{$$}

\def\dd{\displaystyle}

\def\Lt{{{\tilde L}}}
\def\Rt{{{\tilde R}}}

\def\Z{\bf Z}

\newcount \nfor

\def \form {\global \advance \nfor by 1 \eqno(2.\the\nfor)}

\tenrm\baselineskip=13pt
\leftline{\tenbf 1. Introduction}
\vglue0.4cm
The standard $SU(2)\otimes U(1)$ gauge theory of the electroweak
interactions is in good agreement with the current experimental
data, apart the 2 - 3 $\sigma$
discrepancies in $R_b$ and $A_{LR}$.
 Nevertheless there is no yet evidence for the mechanism
which is responsible for the breakdown of the symmetry to the
$U(1)$ electromagnetic.
It is usually assumed that the breaking of the electroweak
symmetry is due to the vacuum expectation value of some elementary scalar.

In this talk I would like to discuss a different option,
a dynamical breaking of the electroweak symmetry: some new interaction
induces a breaking at a scale $\Lambda$ of order 1 TeV.
Effective theories  can be built on the basis of the low energy symmetry
properties. We can build the low energy theory describing goldstones
using the classical technique of Callan, Coleman, Wess and Zumino
(CCWZ)~$^1$, treating the pseudoscalars as the goldstone bosons of a
 spontaneously broken symmetry $G$ to a subgroup $H$.
In the simplest example a chiral symmetry $G=SU(2)_L\otimes SU(2)_R$
is broken to the diagonal subgroup $H=SU(2)_{L+R}$, producing three
Goldstone bosons, which become via the Higgs mechanism the longitudinal
degrees of freedom of $W$ and $Z$.
In general such a theory can contain also new resonances, like
the $\rho$ in QCD.

To build the effective low energy theory describing
Goldstones and vectors, one can use the CCWZ non linear representations
of a chiral symmetry $G$ and considering (\`a la Weinberg~$^2$)
the $\rho$ as the gauge field of
 the unbroken symmetry group $H$. This theory is not renormalizable
in the standard sense.
We can order the terms in the lagrangian
in an energy expansion according the number of derivatives
and truncate at some finite order. The higher order terms
will be proportional to the inverse power of the parameter
$\Lambda$.

In a completely equivalent way one can use
the {\it hidden gauge symmetry} approach. Theories with
non linearly realized symmetry $G\to H$ can be linearly realized by
enlarging the gauge symmetry $G$ to $G\otimes H^\prime\to
H_D=diag(H\otimes H^\prime)$. $H^\prime$ is a local gauge group and the $\rho$
is the gauge field associated to $H^\prime$~$^{3,4}$.

The BESS (Breaking Electroweak Symmetry
Strongly) model was built in this way, using $G=SU(2)_L\otimes SU(2)_R$,
$H=SU(2)_V$~$^5$, and considering the
gauging of the $SU(2)_W\otimes U(1)_Y$.
This model is an effective lagrangian parametrization
of the electroweak symmetry breaking. A new triplet of
vector bosons, mixed with $W$ and $Z$, is present. The parameters
of the BESS model are the mass $M_V$ of these new bosons, their self coupling
$\gs$ and a third parameter $b$ whose strength characterizes the
direct coupling of $V$ to the fermions.
The new charged vector bosons can be studied in the channel
$W^\pm Z\to l^\pm \nu 2l$ at LHC, after their Drell-Yan
production from to the initial quarks,  up to masses of the
order of 2 TeV~$^6$.

At the previous LCWs in Saariselk\"a~$^7$
and in Waikoloa~$^8$ I discussed how future
$e^+e^-$ colliders could restrict the parameter space of the
BESS model.

In principle such a theory can also include axial vector resonances~$^9$,
like the $a_1$ of QCD, or can have a larger symmetry like
$SU(8)\otimes SU(8)$ ~$^{10}$.

In this talk I will present the
results of a new
phenomenological analysis
on a particular version of these models, based on a chiral
$SU(2)_L\otimes SU(2)_R$, containing new vector and axial-vector particles
degenerate in mass (degenerate BESS)~$^{11}$.
This particular choice of the parameters corresponds  to an
enlarged symmetry, and implies that
leading contribution (in the large $M_V$ expansion) to the $\eps_{1,2,3}$
(or $S,T,U$) parameters is zero; therefore the model is not much
 constrained by
the existing data. In degenerate BESS relatively light resonances are
compatible with the electroweak data, as given by LEP and Tevatron.
\vglue0.6cm
\tenrm\baselineskip=13pt
\leftline{\tenbf 2. The model}
\vglue0.4cm

Let me firstly recall how the most general lagrangian up two derivatives
for the linearly realized
$SU(2)_L\otimes SU(2)_R\otimes SU(2)_V\to SU(2)$ symmetry is built. The
coordinates of the manifold $G/H=SU(2)_L\otimes SU(2)_R/SU(2)$
are substituted by a group element $g=(L,R)\in G$.
The Goldstones bosons are represented by two unitary matrices $L$ and $R$
whose transformations are
$$
L\to g_L Lh\quad R\to g_R R h
$$
with $g_{L,R}\in SU(2)_{L,R}$ and $h\in SU(2)_{V}$. Using these fields
one can reconstruct the field $U=L R^\dagger$ which transforms
as $U\to g_L U g_R^\dagger$ and describes the usual field of
$SU(2)_L\otimes SU(2)_R/SU(2)$. We introduce also a gauge field
$V_\mu={\dd i\over 2} \gs{ \dd \tau_i\over 2} V^i_\mu$
 in $Lie SU(2)_V$ and build the covariant derivatives
$D_\mu L=\de_\mu L -LV_\mu$, $D_\mu R=\de_\mu R -RV_\mu$.
The leading terms in the effective lagrangian invariant with respect
to $SU(2)_L\otimes SU(2)_R$ and $L\leftrightarrow R$ transformation
are given
by
$$
L_{eff}=-{{v^2}\over 4} \left [ Tr (L^\dagger D_\mu L -
R^\dagger D_\mu R)^2+ \alpha Tr (L^\dagger D_\mu L +
R^\dagger D_\mu R)^2\right ]
+...
$$
where $v$ and $\alpha $ are arbitrary parameter.
Going into the unitary gauge
$L=R^\dagger= \exp [i \tau_i \pi_i/(2v)]$
one gets an effective lagrangian describing goldstones
and massive vector mesons with $M_V=v/2 \gs\sqrt{\alpha}$.

After the $SU(2)_W\otimes U(1)_Y$ gauging and the identification
$v^2=1/(\sqrt 2 G_F)$, $G_F$ being the Fermi constant, one
get the BESS model ~$^5$.

This procedure can be extended to include also
axial vector resonances. Let $G=SU(2)_L\otimes SU(2)_R$ and
$H'=SU(2)_L\otimes SU(2)_R$.
 The nine Goldstone bosons resulting from the spontaneous breaking
of $G'=G\otimes H'$ to $H_D$, can be described by three independent
$SU(2)$ elements: $L$, $R$ and $M$, with the
following transformations properties
\be
L'= g_L L h_L,\quad R'= g_R R h_R,\quad M'= h_R^\dagger M h_L
\form\ee
with $g_{L,R}\in SU(2)_{L,R}\subset G$ and $h_{L,R}\in H'$.
Moreover we shall require the
invariance under the discrete left-right transformation
$
P:\quad L\leftrightarrow R,\quad M\leftrightarrow M^\dagger
$
which combined with the usual space inversion allows to build the
parity transformation on the fields.
If we ignore the transformations of eq.(2.1), the largest possible global
symmetry of the low-energy theory is given by the requirement of maintaining
for the transformed variables $L'$, $R'$ and $M'$ the character of $SU(2)$
elements, or $G_{max}=[SU(2)\otimes SU(2)]^3$,
consisting of three independent $SU(2)\otimes SU(2)$ factors, acting on each of
the three variables separately. As we shall see, it happens
that, for specific choices of the
parameters of the theory, the symmetry $G'$ gets enlarged to $G_{max}$.

The most general $G'\otimes P$ invariant lagrangian is given by ~$^9$
\be
{L}_G=-\frac{v^2}{4} [a_1 I_1 + a_2 I_2 + a_3 I_3 + a_4 I_4]\form
\ee
plus the kinetic terms ${L}_{kin}$.  The four invariant
terms $I_i$ ($i=1,...4$) are
given by:
\be
I_1=tr[(V_0-V_1-V_2)^2]\quad
I_2=tr[(V_0+V_2)^2]\quad
I_3=tr[(V_0-V_2)^2]\quad
I_4=tr[V_1^2]
\ee
where
\be
V_0^\mu=L^\dagger D^\mu L\quad
V_1^\mu=M^\dagger D^\mu M\quad
V_2^\mu=M^\dagger(R^\dagger D^\mu R)M
\ee
and the covariant derivatives are
\be
D_\mu L=\partial_\mu L -L \lmu\quad
D_\mu R=\partial_\mu R -R \rmu
\ee

\be
D_\mu M=\partial_\mu M -M \lmu+\rmu M
\ee
where $\lmu (\rmu)$ are gauge fields of $SU(2)_{L(R)}\subset H^\prime$
(instead of working with
vector and axial-vector we work with these left and right combinations).

The kinetic terms are given by
\be
{L}_{kin}=\frac{1}{\gs^2} tr[F_{\mu\nu}({\tilde L})]^2+
	 \frac{1}{\gs^2}  tr[F_{\mu\nu}({\tilde R})]^2
\ee
where $\gs$ is the gauge coupling constant for the gauge fields $\lmu$ and
$\rmu$,
and
$F_{\mu\nu}({\tilde L})$, $F_{\mu\nu}({\tilde R})$ are the usual field tensors.

The model I will discuss is characterized by the following choice
of parameters
$a_4=0$, $a_2=a_3$~$^{11}$.
 In order to discuss the symmetry properties it is useful to
observe that the invariant $I_1$ could be re-written as
$
I_1=-tr(\partial_\mu U^\dagger \partial^\mu U)
$
with $U=L M^\dagger R^\dagger$
and the lagrangian as
\be
{L}_G=\frac{v^2}{4}\{a_1~ tr(\partial_\mu U^\dagger \partial^\mu U) +
			 2~a_2~ [tr(D_\mu L^\dagger D^\mu L)+
			  tr(D_\mu R^\dagger D^\mu R)]\}
\form\ee
Each of the three terms in the above expressions
is invariant under an independent $SU(2)\otimes SU(2)$
group
\be
U'=\omega_L U \omega_R^\dagger,\quad L'= g_L L h_L,\quad R'= g_R R h_R
\ee
The overall symmetry is $G_{max}=[SU(2)\otimes SU(2)]^3$, with a part
$H'$ realized as a gauge symmetry.
With the particular choice $a_4=0$, $a_3=a_2$, as we see from eq.(2.3),
the mixing between $\lmu$ and $\rmu$ is vanishing, and the new states are
degenerate in mass.
Moreover, as it follows from eq.(2.3), the longitudinal modes
of the fields are entirely
provided by the would-be Goldstone bosons in $L$ and $R$. This means
that the pseudoscalar particles remaining as physical states in the
low-energy spectrum are those associated to $U$. They in turn can
provide the longitudinal components to the $W$ and $Z$ particles,
in an effective description of the electroweak breaking sector.

The peculiar feature of degenerate BESS is that the new bosons are
not coupled to those Goldstone bosons which are absorbed to give
mass to $W^\pm$ and $Z$. As a consequence the channels $W_L Z_L$ and
$W_L W_L$ are not strongly enhanced as it usually happens in models with a
strongly interacting symmetry breaking sector and this implies larger branching
ratios of the new resonances into fermion pairs.

The coupling of the model to the electroweak
$SU(2)_W\otimes U(1)_Y$ gauge fields is obtained
via the minimal substitution
$$
D_\mu L \to D_\mu L+ {\Wt}_\mu L\quad
D_\mu R \to D_\mu R+ {\Yt}_\mu R\quad
D_\mu M \to D_\mu M
$$
where
$$
\eqalign{
\Wt_\mu&=i g {\tilde W}_\mu ^a\f{\tau^a}{2}\quad
\Yt_\mu=i\gp
 {\tilde Y}_\mu\f{\tau^3}{2}\cr
\Lt_\mu&=i\f{\gs}{\sqrt{2}} {\tilde L}_\mu ^a\f{\tau^a}{2}\quad
\Rt_\mu=i\f{\gs}{\sqrt{2}} {\tilde R}_\mu ^a\f{\tau^a}{2}}
$$
with $g$, $\gp$ the $SU(2)_W\otimes U(1)_Y$
gauge coupling constant and $\tau^a$ the Pauli matrices.

By introducing the canonical kinetic terms for $W_\mu^a$ and $Y_\mu$
and going into the unitary gauge we get
$$
\eqalign{
\L=&-\f{v^2}{4}\Big[ a_1 \tr(\Wt_\mu-\Yt_\mu)^2
+2 a_2 \tr(\Wt_\mu-{\Lt}_\mu)^2
+2 a_2 \tr(\Yt_\mu-\Rt_\mu)^2\Big]\cr
&+\L^{kin}(\Wt,\Yt,\Lt,\Rt)}\form
$$


We have used tilded quantities to reserve untilded variables
for mass eigenstates.

The standard model (SM)
relations are obtained in the limit $\gs \gg {g},
{g}'$. Actually,
for a very large $\gs$, the kinetic terms for the fields $\lmu$ and $\rmu$
drop out, and ${\cal L}$ reduces to the first term in eq.(2.4).
This term reproduces precisely the mass term
for the ordinary gauge vector bosons in the SM, provided we
assume  $a_1=1$.
Finally let us consider the fermions of the SM and denote them by $\psi_L$
and $\psi_R$. They couple to $\Lt$ and $\Rt$ via the mixing with the
standard $\Wt$ and $\Yt$:
$$
\eqalign{
\L_{fermion} &= \overline{\psi}_L i \gamma^\mu\Big(\dmu+
i g {\tilde W}_\mu ^a\f{\tau^a}{2}+
		      \f{i}{2}\gpt(B-L){\tilde Y}_\mu\Big){\psi}_L\cr
     &+\overline{\psi}_R i \gamma^\mu\Big(\dmu+
i\gp {\tilde Y}_\mu\f{\tau^3}{2}+\f{i}{2}\gpt
		      (B-L) \Yt_\mu\Big){ \psi}_R}
$$
where $B(L)$ is the baryon (lepton) number, and
$\psi =\left(\psi_u,\psi_d\right)$.


By separating  the charged and the neutral gauge  bosons
the quadratic lagrangian is given by:

$$
\eqalign{
{\cal L}^{(2)} &= \frac{v^2}{4}[(1+2 a_2)g^2 \tW_\mu^+ \tW^{\mu -}+
		      a_2 \gs^2 (\tL_\mu^+ \tL^{\mu -}+\tR_\mu^+
		  \tR^{\mu -})\cr
		& -\sqrt{2}a_2g \gs (\tW_\mu^+ \tL^{\mu -}+\tW_\mu^-
		\tL^{\mu +})]\cr
&+ \frac{v^2}{8}[(1+2 a_2) (g^2 \tW_3^2+\gptd \tY^2)+
		      a_2 \gs^2 (\tL_3^2+\tR_3^2)\cr
		 &- 2 g \gpt \tW_{3\mu}\tY^\mu
		-2 \sqrt{2}a_2\gs (g \tW_3 \tL_3^\mu +\gp \tY_\mu
		 \tR_3^\mu)]}
\form$$

Therefore the   $R^\pm$ fields are unmixed and their mass can be easily
read: $M_{R^{\pm}}\equiv M=v \gs  \sqrt  a_2/2$.
We will parametrize the model by using, in addition to
the SM parameters, $M$ and $g/\gs$.

\vglue13pt
\centerline{
\epsfxsize=8truecm
\epsffile[78 263 489 700]{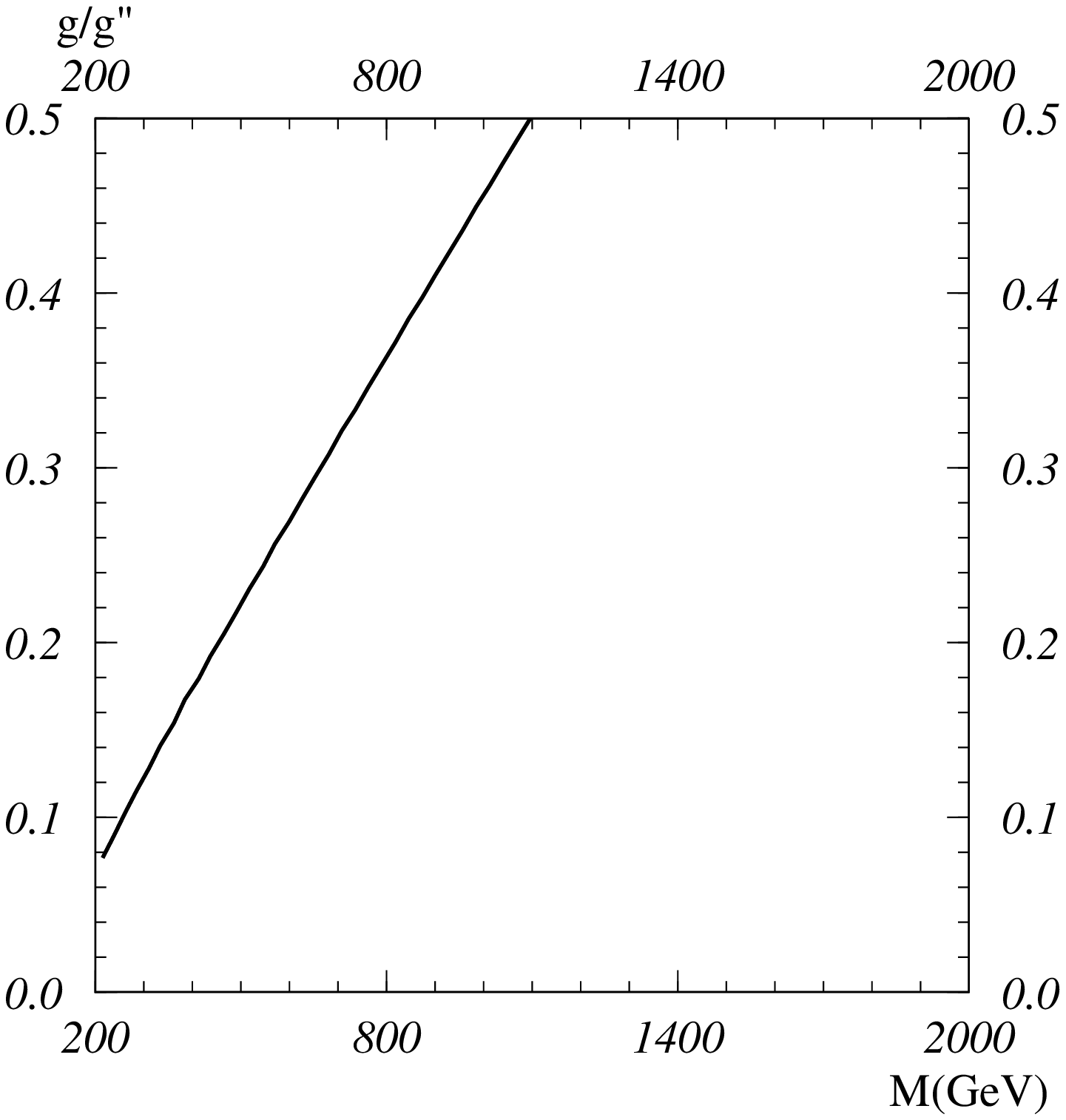}}
\vglue13pt
{\eightrm Fig.~1. 90\% C.L. contour on the plane ($M$, $g/g''$) obtained by
comparing the values of the $\epsilon$ parameters
from the model with the experimental data from LEP.
 The allowed region is below the curve.}
\smallskip
Eigenvalues and eigenvectors
for the remaining fields can be found in ~$^{11}$. As  already
said the heavy fields have all the degenerate mass $M$ in the large
$\gs$ limit. By using eq.(2.5) one can show that at the leading
order in $q^2/M^2$  the contribution of the model to all
$\eps$ parameters~$^{12}$ is equal to zero~$^{11}$.
 This is due to the fact that in the
$M\to\infty$ limit, this model decouples.
We can perform the low-energy limit at the next-to-leading order and study
the virtual effects of the heavy particles. Working at the first order in
$1/\gs^2$ we get
$\epsilon_1=-(\c^4+\s^4)/(\c^2)~ X$,
$\epsilon_2=-\c^2~ X$,
$\epsilon_3=-X$
with $X=2({M_Z^2}/{M^2})( g/\gs)^2$.
All these deviations are of order $X$ which contains a double
suppression factor $M_Z^2/M^2$ and $(g/\gs)^2$.
The sum of the SM contributions, functions of the top and Higgs masses,
and of these deviations has to be compared with the experimental
values for the $\epsilon$ parameters, determined from the all
available LEP data and the $M_W$ measurement at Tevatron ~$^{13}$:
$\epsilon_1=(3.8 \pm 1.5)\cdot 10^{-3}$,
$\epsilon_2=(-6.4\pm 4.2)\cdot 10^{-3}$,
$\epsilon_3=(4.6\pm 1.5)\cdot 10^{-3}$.
Taking into account the SM values
$(\epsilon_1)_{SM}=4.4\cdot 10^{-3}$,
$(\epsilon_2)_{SM}=-7.1\cdot 10^{-3}$,
$(\epsilon_3)_{SM}=6.5\cdot 10^{-3}$
 for $m_{top}=180~GeV$ and
$m_H=1000~GeV$, we find, from the combinations of the previous
experimental results, the $90\%$ C.L. limit on $g/\gs$ versus
the mass M given by the solid line in Fig.1. The allowed
region is the one below the continuous line.

\tenrm\baselineskip=13pt
\vglue0.6cm\leftline{\tenbf 3. Degenerate BESS at $e^+e^-$ future colliders}
\vglue0.4cm
In this section I will discuss
the sensitivity of the model at LEP2 and future $e^+e^-$
linear colliders, for different options of total centre of mass energies and
luminosities.

Cross-sections and asymmetries for the channel
$e^+e^-\rightarrow f^+f^-$ and $e^+e^-\rightarrow W^+W^-$ in the Standard
Model and in the degenerate BESS model at tree level have been studied
$^{11}$.
The BESS states relevant for the
analysis at $e^+e^-$ colliders are $L_3$ and $R_3$. Their coupling
to fermions can be found in ~$^{11}$.
I will not
consider the direct production of $R_3$ and $L_3$ from $e^+e^-$,
but rather their indirect effects in the $e^+e^-\rightarrow f^+f^-$ and
$e^+e^-\rightarrow W^+W^-$ cross-sections.
In the fermion channel the study is based on the following observables:
the total hadronic ($\mu^+\mu^-$)
cross-sections $\sigma^h$ ($\sigma^{\mu}$),
the forward-backward and left-right
asymmetries  $A_{FB}^{e^+e^- \to \mu^+ \mu^-}$,
$A_{FB}^{e^+e^- \to {\bar b} b}$,
$A_{LR}^{e^+e^- \to \mu^+ \mu^-}$,
$A_{LR}^{e^+e^- \to h}$ and  $A_{LR}^{e^+e^- \to {\bar b} b}$.
At LEP2 we can add to the previous observables the $W$ mass measurement.
The result of this
analysis
shows that LEP2 will not improve considerably the existing limits $^{14}$.

\vglue13pt
\centerline{
\epsfxsize=8truecm
\epsffile[78 263 489 700]{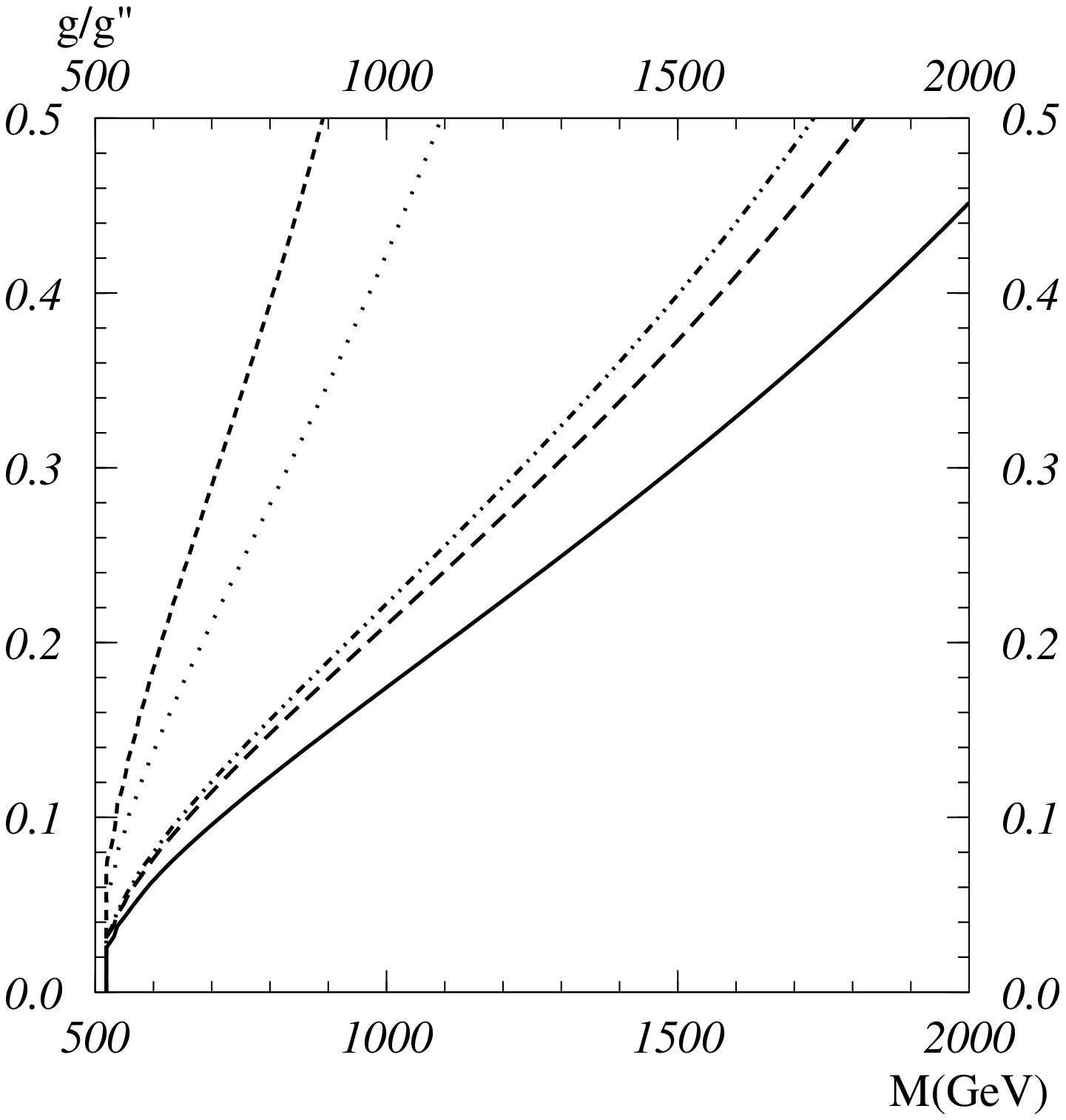}}
\vglue13pt
{\eightrm Fig.~2.
90\% C.L. contour on the plane
($M$, $g/g''$) from $e^+e^-$ at $\sqrt{s}=500~GeV$ with
an integrated luminosity of $20 fb^{-1}$ from unpolarized observables.
Allowed regions are below the curves.
(Dashed-dotted $\sigma^h$, dashed $\sigma^\mu$, dotted
$A_{FB}^\mu$, the uppermost dashed $A_{FB}^b$,
continuous all combined).}
\smallskip
To improve these limits it is necessary to consider higher energy colliders.
Two options for a high energy $e^+e^-$ collider have been
studied:
$\sqrt{s}=500~GeV$ ($\sqrt{s}=1~TeV$)
with an integrated luminosity of $20 fb^{-1}$ ($80 fb^{-1}$).

In Fig. 2 we present the 90\% C.L. contour on the plane
($M$, $g/g''$) from $e^+e^-$ at $\sqrt{s}=500~GeV$ with
an integrated luminosity of $20 fb^{-1}$ for various
observables. The dashed-dotted line
represents the limit from $\sigma^h$ with an assumed relative error of 2\%;
 the dashed
line near to the preceeding one is $\sigma^\mu$ (relative
error 1.3\%), the dotted line
is $A_{FB}^\mu$  (error 0.5\%) and the uppermost dashed line is $A_{FB}^b$
(error
0.9\%).

As it is evident more stringent bounds come from the
cross-section measurements. Asymmetries give less restrictive bounds due to
a compensation between the $L_3$ and $R_3$ exchange.
By combining all the deviations in the previously considered
observables we get the limit shown by the continuous line.

\vglue13pt
\centerline{
\epsfxsize=8truecm
\epsffile[78 263 489 700]{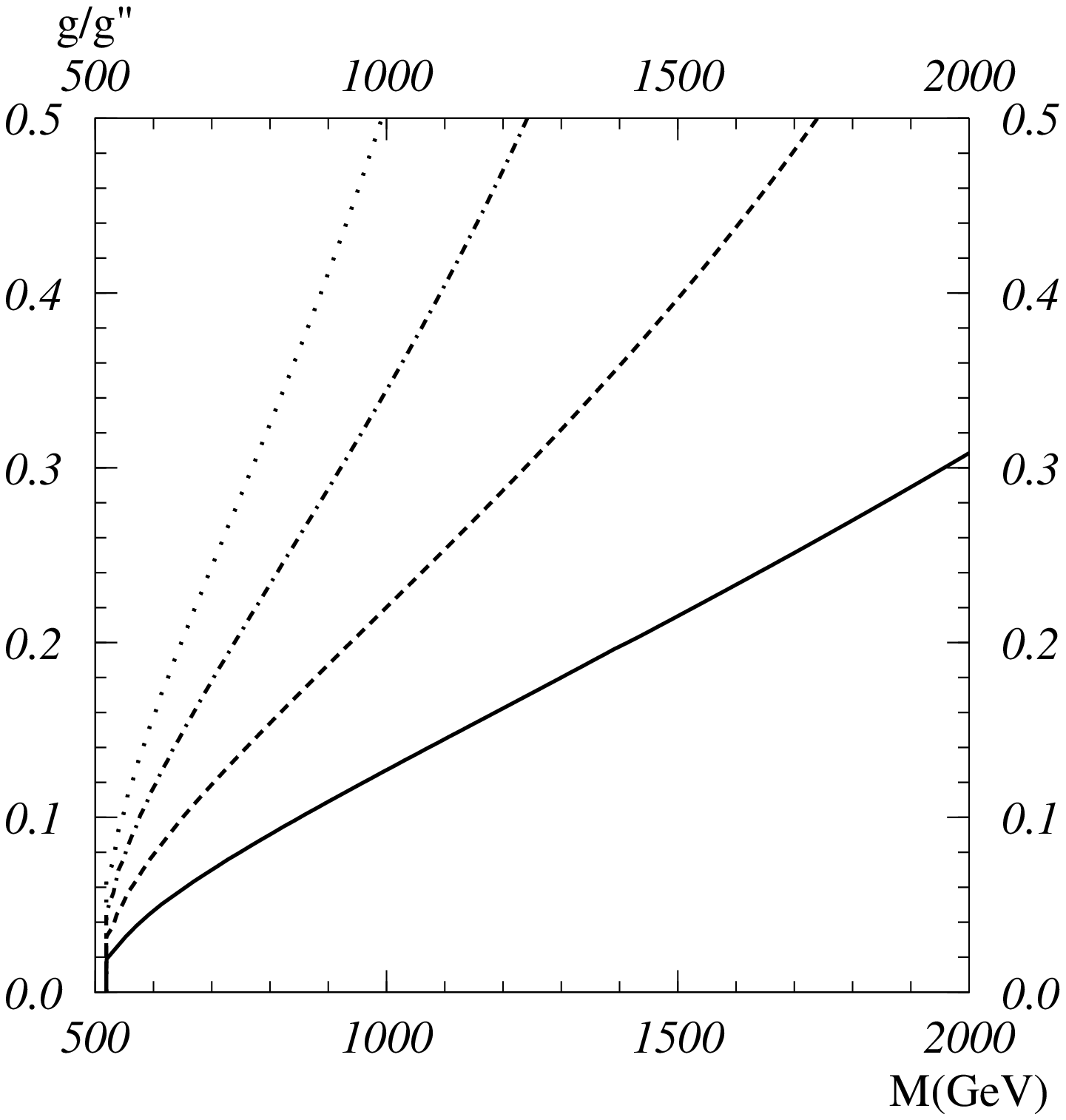}}
\vglue13pt
{\eightrm Fig.~3.
90\% C.L. contour on the plane
($M$, $g/g''$) from $e^+e^-$ at $\sqrt{s}=500~GeV$ with
an integrated luminosity of $20 fb^{-1}$ from polarized observables.
Allowed regions are below the curves.
(Dashed-dotted $A_{LR}^\mu$, dashed $A_{LR}^h$,
dotted $A_{LR}^b$, continuous all unpolarized and polarized combined).}
\smallskip

Polarized electron beams
allow to get further limit in the parameter space as shown in Fig. 3.
We neglect the error on the measurement of the polarization and use a
polarization value equal to 0.5.
The dashed-dotted line represents the limit from $A_{LR}^\mu$ (error  0.6\%),
the dashed line from $A_{LR}^h$
(error 0.4\%), the dotted line from $A_{LR}^b$
 (error 1.1\%).
Combining all the polarized and unpolarized beam observables we get the
bound shown by the continuous line. In conclusion
a substantial improvement with respect to the LEP bounds,
even without polarized beams is obtained.

\vglue13pt
\centerline{
\epsfxsize=8truecm
\epsffile[78 263 489 700]{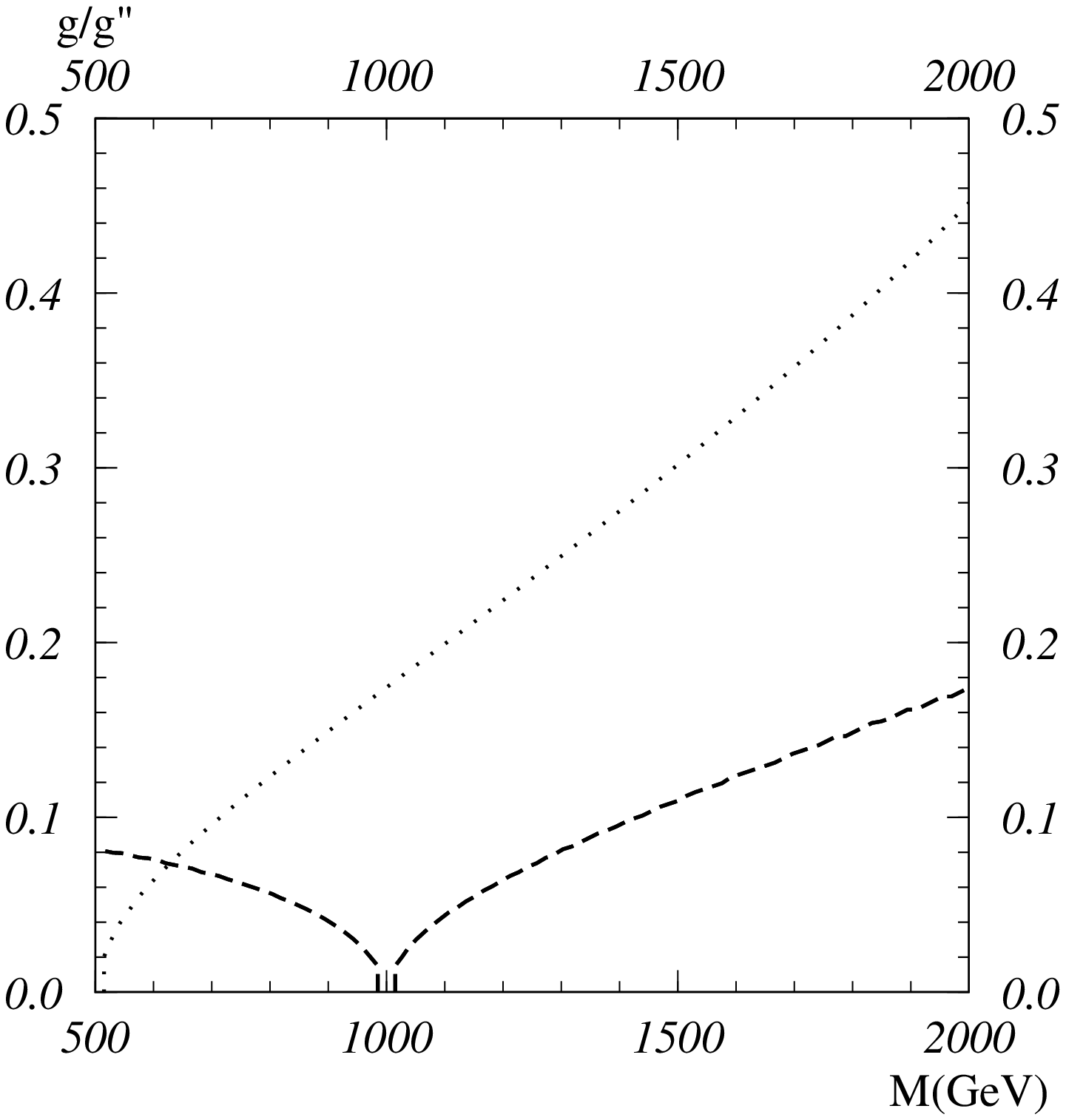}}
\vglue13pt
{\eightrm Fig.~4. 90\% C.L. contour on the plane ($M$, $g/g''$) from
$e^+e^-$ at $\sqrt{s}=500~GeV$ with an
integrated luminosity of $20 fb^{-1}$ and $\sqrt{s}=1000~GeV$
with an integrated luminosity of $80 fb^{-1}$.
Allowed regions are below the curves.}
\smallskip

In Fig. 4 a combined picture of the
 90\% C.L. contours on the plane ($M$, $g/g''$) from $e^+e^-$
at two values of $\sqrt{s}$ is shown.  The dotted
line represents the limit from the combined unpolarized observables at
$\sqrt{s}=500~GeV$ with an integrated luminosity of $20 fb^{-1}$; the
dashed line is the limit from the combined unpolarized observables
at $\sqrt{s}=1000~GeV$ with an integrated luminosity of $80 fb^{-1}$.
As expected increasing the energy of the collider and rescaling the
integrated luminosity result in stronger bounds on the
parameter space.

The $WW$ final state, considering the observables
given in $^{11}$ has been also studied. However
the new channel does  not modify the strong limits
obtained using the fermion final state. This is  because
the degenerate model has no strong
enhancement of the $WW$ channel, present in the usual
strong electroweak models.


\vglue0.6cm
\leftline{\tenbf 4. Acknowledgements}
\vglue0.4cm
I would like to thank
R. Casalbuoni, A. Deandrea, S. De Curtis, R. Gatto
and M. Grazzini, for the enjoyable collaboration which
led to the results presented in this talk.

\vglue0.6cm
\leftline{\tenbf 5. References}
\vglue0.4cm
\medskip

\itemitem{1.} Coleman, J. Wess and B. Zumino, {\it Phys. Rev. }{\bf 177}
 (1969) 2239;
C. G. Callan, S. Coleman, J. Wess and B. Zumino,
{\it Phys. Rev. }{\bf 177} (1969) 2247.

\itemitem{2.} S. Weinberg, {\it \it Phys. Rev. } {\bf 166} (1968) 1568.

\itemitem{3.} A.P. Balachandran, A. Stern and G. Trahern, {\it
Phys. Rev.} {\bf
D19 } (1979) 2416.

\itemitem{4.} M. Bando, T. Kugo, S. Uehara, K. Yamawaki and T.
Yanagida,  {\it Phys. Rev. Lett.} {\bf 54} (1985) 1215.

\itemitem{5.}
R. Casalbuoni, S. De Curtis, D. Dominici and R. Gatto,
      {\it Phys. Lett. } {\bf B155}  (1985) 95; and
      {\it  Nucl. Phys.} {\bf B282} (1987) 235.

\itemitem{6.}R. Casalbuoni, P. Chiappetta, S. De Curtis,
F. Feruglio, R. Gatto,
B. Mele and J. Terron, {\it Phys. Lett.} {\bf B249} (1991) 130.
I. Josa, T. Rodrigo and F. Pauss,
   CERN 90-10, volume II, p. 796, Proceedings
of Large Hadron Collider Workshop, Aachen, 4-9 October 1990, Eds. G. Jarlskog.

\itemitem{7.} D.Dominici, in {\it Physics and
Experiments with Linear Colliders}, eds. R. Orava, P. Eerola and
M. Nordberg (World Scientific,
Singapore, 1992), p. 509.

\itemitem{8.} R. Casalbuoni, S. De Curtis, D. Dominici, P. Chiappetta,
A. Deandrea and R. Gatto, in {\it Physics and
Experiments with Linear $e^+e^-$ Colliders\/}, eds. F. A. Harris,
S. L. Olsen, S. Pakvasa and X. Tata (World Scientific,
Singapore, 1993), p. 887.

\itemitem{9.}
R. Casalbuoni, S. De Curtis, D. Dominici F. Feruglio and R. Gatto,
       Int. Jour. Mod. Phys. {\bf A4}  (1989) 1065.

\itemitem{10.}
R. Casalbuoni, A. Deandrea, S. De Curtis, N. Di Bartolomeo, D. Dominici,
F. Feruglio and R. Gatto, Nucl. Phys. {\bf B409} (1993) 257.

\itemitem{11.}
R. Casalbuoni, A. Deandrea, S. De Curtis, D. Dominici,  R. Gatto
and M. Grazzini, UGVA-DPT 1995/10-906,  hep-ph/9510431.

\itemitem{12.}
G. Altarelli and R. Barbieri, Phys. Lett. {\bf B253} (1991) 161;
G. Altarelli, R. Barbieri and S. Jadach, Nucl. Phys. {\bf B369} (1992) 3;
G. Altarelli, R. Barbieri and F. Caravaglios, Nucl. Phys. {\bf B405} (1993) 3.

\itemitem{13.}
F. Caravaglios, talk at the EPS conference on Particle Physics, Bruxelles,
July 1995.

\itemitem{14.}
R. Casalbuoni, A. Deandrea, S. De Curtis, D. Dominici,  R. Gatto
and M. Grazzini, Proceedings of the CERN Workshop on Physics
at Lep2, CERN Yellow Report, to appear in 1996.

\medskip

\bye